\documentclass[reprint,aps,prb,superscriptaddress,twocolumn,floatfix]{revtex4-2}

\usepackage{graphics,graphicx}
\usepackage{bm}
\usepackage[abs]{overpic}
\usepackage{amsmath,amssymb,amsfonts}
\usepackage[breaklinks=true,colorlinks,citecolor=blue,linkcolor=blue,urlcolor=blue]{hyperref}

\usepackage[abs]{overpic}

\usepackage{mathrsfs}

\usepackage{stfloats}

\newcommand{\be}{\begin{eqnarray}}
\newcommand{\ee}{\end{eqnarray}}
\newcommand{\nn}{\nonumber\\}

\begin{document}

\title{Ultrafast electronic heat dissipation through surface-to-bulk Coulomb coupling in quantum materials.}
\author{Alessandro Principi}
\email{alessandro.principi@manchester.ac.uk}
\affiliation{\noindent Department of Physics and Astronomy, University of Manchester, Manchester M13 9PL, UK}
\author{Klaas-Jan Tielrooij}
\affiliation{\noindent Catalan Institute of Nanoscience and Nanotechnology (ICN2), BIST \& CSIC,
Campus UAB, 08193 Bellaterra, Barcelona, Spain}

\begin{abstract}
The timescale of electronic cooling is an important parameter controlling the performance of devices based on quantum materials for optoelectronic, thermoelectric and thermal management applications.  In most conventional materials, cooling proceeds via the emission of phonons, a relatively slow process that can bottleneck the carrier relaxation dynamics, thus degrading the device performance. Here we present the theory of near-field radiative heat transfer, that occurs when a two-dimensional electron system is coupled via the non-retarded Coulomb interaction to a three-dimensional bulk that can behave as a very efficient electronic heat sink. We apply our theory to study the cooling dynamics of surface states of three dimensional topological insulators, and of graphene in proximity to small-gap bulk materials. The ``Coulomb cooling'' we introduce is alternative to the conventional phonon-mediated cooling, can be very efficient and dominate the cooling dynamics under certain circumstances. We show that this cooling mechanism can lead to a sub-picosecond time scale, significantly faster than the cooling dynamics normally observed in Dirac materials. 
\end{abstract}

\maketitle
 
\section{Introduction}
Two-dimensional (2D) Dirac materials~\cite{castroneto_rmp_2009,Hasan_rmp_2010,Hasan_annrev_2011,Wehling_advphys_2014} have been extensively studied in the last two decades for the diverse range of intriguing properties they harbor, which could in turn enable a wealth of novel practical applications~\cite{fu_prl_2007,burkov_prl_2010,Pesin_natmat_2012,Lupi_appmat_2020}. This is the case of graphene~\cite{castroneto_rmp_2009,Kotov_rmp_2012,dassarma_rmp_2011,Massicotte_Nanoscale_2021}, arguably the most-studied Dirac material, which continues to attract significant interest because of its potential for electronic and optoelectronic applications~\cite{Massicotte_Nanoscale_2021}. 
However, graphene is just an instance of a broad family of systems which includes the 2D surface states of three-dimensional (3D) topological insulator~\cite{Hasan_rmp_2010,Hasan_annrev_2011,Qi_rmp_2011}. Their inverted bulk band structure allows states to localize at the surface~\cite{Kane_prl_2005,fu_prl_2007,Fu_prb_2007,Teo_prb_2008,Yan_rpp_2012}. In ideal topological insulators, surface states do not hybridize with bulk ones, and are topologically protected against any perturbation that preserves the symmetries of the bulk~\cite{Hasan_rmp_2010,Hasan_annrev_2011,Qi_rmp_2011}.

Thanks to the coupling between kinetic momentum and spin, the electronic surface states of a 3D topological insulator present a physics potentially richer than graphene, and promise application to a diverse range of fields, including spintronics~\cite{burkov_prl_2010,Pesin_natmat_2012}, optoelectronics, and photonics~\cite{Politano_aplmat_2017,Pandey_scirep_2021}. In a recent experimental work~\cite{Kovalev_npjqm_2021}, the cooling dynamics of surface electrons of bismuth and antimony chalcogenides was studied with pump-probe techniques. Surprisingly, electronic heat relaxation faster than that of bulk carriers was observed. The observed bulk and surface heat-decay rates differ by about an order of magnitude, while the environment experienced by their electrons is the same. Interestingly, a decay rate of a few hundred femtoseconds was obtained under rather strong photoexcitation with a fluence on the order of $100 \mu{\rm J/cm}^2$. In comparison, when graphene is excited with a similar fluence, cooling is rather slow - several picoseconds - due to a phonon bottleneck effect~\cite{Deinert_acsnano_2021}. 
 
So why does the cooling of the electrons of surface states of topological insulators occur so much faster than their bulk counterpart or than graphene?
In graphene, the cooling of photoexcited electrons is ultimately limited by the emission of intrinsic optical phonons of the material~\cite{Pogna_ACSnano_2021} or of the encapsulant~\cite{Principi_prl_2017,Tielrooij_natnano_2018}. Such phonon emission mechanisms fail to account for the starkly different relaxation dynamics of surface and bulk electrons of topological insulators, unless, one would postulate different electron-phonon couplings or a reduced phase space for phonon emission by surface states. 

\begin{figure}[t]
\begin{center}
\begin{tabular}{c}
\includegraphics[width=0.98\columnwidth]{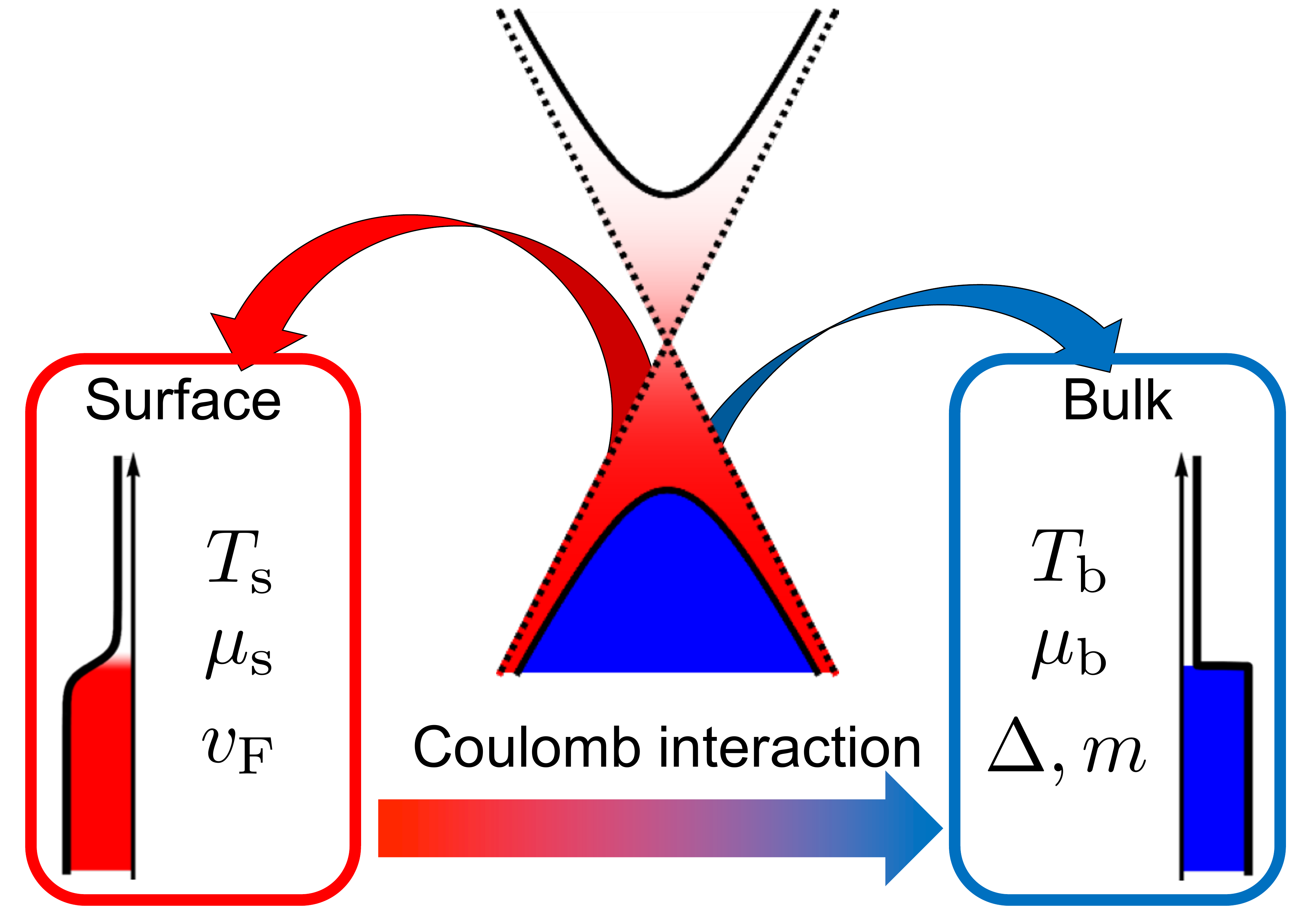}
\end{tabular}
\end{center}
\caption{
Schematics of Coulomb cooling of hot surface Dirac fermions into bulk states. Dirac fermions have a constant Fermi velocity $v_{\rm F}$, and exhibit a hot thermal distribution characterized by temperature and chemical potential $T_{\rm s}$ and $\mu_{\rm s}$, respectively. Similarly, bulk states are described in terms of the effective mass $m$ and band gap $2\Delta$. The Fermi distribution of bulk states has chemical potential $\mu_{\rm b}$ (set to zero in later calculations) and temperature $T_{\rm b}< T_{\rm s}$.
}
\label{fig:one}
\end{figure}

In this paper, we explore a different mechanism that does not require fine-tuning material parameters, as it relies on electronic systems exclusively. This alternative mechanisms is based on the notion that, contrary to electrons in conventional graphene devices, the surface electrons of topological insulators are in close proximity to a macroscopic bulk whose particle-hole excitations can occur at energies comparable to the (surface) thermal excitations. Thanks to the near-field radiative coupling between surface and bulk, the latter can act as a heat sink. Being macroscopic, the bulk can absorb large amounts of heat and therefore efficiently cool down the surface electrons. In this picture, heat is dissipated into particle-hole excitations of the bulk via non-retarded Coulomb interactions (see Fig.~\ref{fig:one}). Thus we term this mechanism ``Coulomb cooling''~\cite{Mihnev_natcomm_2015}.

In the remainder of the paper, we consider a system of 2D massless Dirac fermions in proximity to a 3D gapped bulk. The two are coupled only electrostatically via non-retarded Coulomb interactions. This minimal model of a topological insulator equivalently describes a graphene sheet in proximity to a small-gap material. For both cases, using the kinetic equation for surface electrons, we derive an expression for their cooling rate. This is controlled by the convolution of surface and bulk particle-hole excitation spectra. The higher their overlap, the larger the amount of heat transferred per unit time. We show that the cooling rate reaches a maximum for surface temperatures close to half the bulk-band gap. This is interpreted as a resonance between surface electronic transitions, whose typical energy is the thermal one, and interband particle-hole bulk excitations across the band gap. We find a timescale of a few hundred femtoseconds for topological insulators, and even a few tens of femtoseconds for graphene on a small-bandgap semiconductor.

\section{The model}
\label{sect:model}
We model both the surface states of the topological insulator and the electrons in graphene as a gas of massless Dirac fermions~\cite{Hasan_rmp_2010,Hasan_annrev_2011,Qi_rmp_2011,castroneto_rmp_2009,Kotov_rmp_2012,dassarma_rmp_2011,Massicotte_Nanoscale_2021}, {\it i.e.}
\be\label{eq:H_s_def}
{\cal H}_{\rm s} = \hbar v_{\rm F} \sum_{{\bm k},\alpha,\beta} {\hat \psi}^\dagger_{{\bm k},\alpha,{\rm s}} {\bm k} \cdot {\bm \sigma}_{\alpha,\beta} {\hat \psi}_{{\bm k},\beta,{\rm s}}
~, 
\ee
where $v_{\rm F}$ is the Fermi velocity and ${\bm \sigma}$ is a vector of Pauli matrices. These act in the real spin space for topological insulators, and in pseudospin (sublattice) space for graphene. In Eq.~(\ref{eq:H_s_def}), ${\hat \psi}^\dagger_{{\bm k},\alpha,{\rm s}}$ (${\hat \psi}_{{\bm k},\alpha,{\rm s}}$) creates (destroys) a surface particle of two-dimensional wavevector ${\bm k}$ (momentum $\hbar {\bm k}$) and (pseudo)spin projection $\alpha$. The band energy is $\varepsilon^{({\rm s})}_{{\bm k},\lambda} = \lambda \hbar v_{\rm F} |{\bm k}|$, where $\lambda = +$ ($\lambda = -$) denotes the surface conduction (valence) band. [The Hamiltonian~(\ref{eq:H_s_def}) is obtained from the usual one for topological insulators, which features a cross product between momentum and spin operators~\cite{Hasan_rmp_2010}, by re-defining the spin quantization axes.] 
Eq.~(\ref{eq:H_s_def}) is in form identical for graphene and topological insulators. However, their electrons exhibit different degeneracies $N_{\rm f}$. While the surface states of topological insulators are helical~\cite{Hasan_rmp_2010}, thus yielding $N_{\rm f} = 1$, in graphene they feature a full spin-valley degeneracy~\cite{castroneto_rmp_2009}. Thus, for graphene, the number of fermion flavors is $N_{\rm f} = 4$.

We model the bulk as a two-band system confined into the half-space $z>0$ and described by the Hamiltonian
\be
{\cal H}_{\rm b} = \sum_{{\bm k},\eta} \varepsilon^{({\rm b})}_{{\bm k},\eta} {\hat \psi}^\dagger_{{\bm k},\eta,{\rm b}} {\hat \psi}_{{\bm k},\eta,{\rm b}}
~.
\ee
Here, $\varepsilon^{({\rm b})}_{{\bm k},\eta} = \eta\big[\Delta + \hbar^2 |{\bm k}|^2/(2m)\big]$, where $2\Delta$ is the bulk-band gap and $m$ is the bulk-band mass (assumed to be the same for both valence and conduction bands), while ${\hat \psi}^\dagger_{{\bm k},\eta,{\rm b}}$ (${\hat \psi}_{{\bm k},\eta,{\rm b}}$) creates (destroys) a bulk particle of three-dimensional wavevector ${\bm k} = ({\bm k}_\parallel,k_z)$, in band $\eta$. Here, ${\bm k}_\parallel$ and $k_z$ are the wavevectors parallel and perpendicular to the surface $z=0$, $\eta = +$ denotes the conduction band, while $\eta = -$ stands for the valence band. For simplicity, we assume that bulk bands are spherically symmetric and have no spin structure. These approximations do not affect the temperature dynamics on a qualitative level. Assuming specular reflection at the interface $z=0$, the bulk eigenstates acquire the form of standing waves: $\Psi_{{\bm k},\eta,{\rm b}}({\bm r},z) = \sqrt{2/V} e^{i{\bm k}_\parallel\cdot {\bm r}} \sin(k_z z)$, where ${\bm r}$ is a vector along the surface.

We assume that bulk and surface electrons are coupled electrostatically by long-range instantaneous Coulomb interactions~\cite{principi_prb_2018}, whose Hamiltonian is
\be
\label{eq:hee}
{\cal H}_{\rm sb} =
\frac{1}{2} \int_0^\infty dz \sum_{{\bm q}} V_{{\rm sb}}({\bm q},z) {\hat n}_{{\bm q},{\rm s}} {\hat n}_{-{\bm q},{\rm b}}(z)
~.
\ee
Here, ${\bm q}$ is a two-dimensional wavevector along the surface of the topological insulator, while ${\hat n}_{{\bm q},{\rm s}}$ and $ {\hat n}_{-{\bm q},{\rm b}}(z)$ are the 2D-Fourier transforms of the surface and bulk density operators. The precise form of the interaction $V_{{\rm sb}}({\bm q},z)$ is determined by the solution of the associated Poisson (electrostatic) problem, as we proceed to show.

\subsection{The interaction between surface and bulk electrons}
\label{app:electrostatics}
In this section we derive the Coulomb interaction between surface and bulk electrons. To do so, we consider the electrostatic problem of a single charge added to a conducting 2D sheet (the surface states located at $z=0$) placed on top of the topological-insulator bulk which fills the half-space $z>0$. The half-space $z<0$ is instead empty. The single charge is added as a plane wave of wavevector ${\bm q}$ in the surface sheet. In response to this added charge, induced bulk and surface densities are generated. These are named $n_{\rm b}({\bm r},z)$ and  $n_{\rm s}({\bm r})$, respectively.
The resulting Poisson equation (in Gaussian units) is 
\be \label{eq:poisson}
{\bm \nabla}\cdot \big[\epsilon(z) {\bm \nabla} \phi({\bm r},z)\big] = -4\pi e^2 n({\bm r},z)
~,
\ee
where $n({\bm r},z) = n_{\rm b}({\bm r},z) \Theta(z) + n_{\rm s}({\bm r})\delta(z) + e^{-i{\bm q}\cdot {\bm r}} \delta(z)$ is the total electron density, and $\epsilon(z) = \epsilon_{\rm b} \Theta(z) + \epsilon_{\rm vac} \Theta(-z)$ is the dielectric function. Here, $\epsilon_{\rm b}$ is the relative dielectric constant of the topological insulator, which accounts for the screening due to filled bands, while $\epsilon_{\rm vac} = 1$ [$\Theta(z)$ is the Heaviside step function].
The Fourier transform of Eq.~(\ref{eq:poisson}) in the direction parallel to the topological-insulator surface (or the graphene sheet) yields
\be \label{eq:poisson_2}
\partial_z \big[\epsilon(z) \partial_z \phi_{\bm q}(z)\big] - q^2 \epsilon(z) \phi_{\bm q}(z) = -4\pi e^2 n_{\bm q}(z)
~,
\ee
where $n_{\bm q}(z) = n_{{\bm q},{\rm b}}(z) \Theta(z) + n_{{\bm q},{\rm s}}\delta(z)+\delta(z)$. Within linear response, we write $n_{{\bm q},{\rm s}} = \chi_{\rm s}({\bm q},\omega)\phi_{\bm q}(0)$, where $\chi_{\rm s}({\bm q},\omega)$ is the density-density response function (polarizability) of the surface states, and 
\be\label{eq:n_b_LR}
n_{{\bm q},{\rm b}}(z) = \int_0^\infty dz' \chi_{\rm b}({\bm q},\omega,z,z') \phi_{\bm q}(z') \simeq {\bar \chi}_{\rm b}({\bm q},\omega) \phi_{\bm q}(z)
~.
\nn
\ee
Here, $\chi_{\rm b}({\bm q},\omega,z,z') \simeq {\bar \chi}_{\rm b}({\bm q},\omega) \delta(z-z')$ is the density-density response function~\cite{Giuliani_and_Vignale} (polarizability) of bulk states treated within a semi-local approximation. Equations of this section contain the response of the doped bulk, embodied by $\chi_{\rm b}({\bm q},\omega,z,z')$ and ${\bar \chi}_{\rm b}({\bm q},\omega)$, for completeness. However, we leave this function unspecified since, as we show below, it plays no role in the undoped regime when the bulk thermal energy ($k_{\rm B} T_{\rm b}$) is much smaller than the bulk-band gap ($2\Delta$).
Using these expressions, the Poisson equation~(\ref{eq:poisson_2}) can be split into the two half-spaces as
\be \label{eq:poisson_3}
\left\{
\begin{array}{ll}
\epsilon_{\rm b}(\partial_z^2 - q^2)\phi_{\bm q}(z) = - 4\pi e^2 {\bar \chi}_{\rm b}({\bm q},\omega) \phi_{\bm q}(z) & {\rm if}~ z>0
\vspace{0.2cm}\\
(\partial_z^2 - q^2)\phi_{\bm q}(z) = 0 & {\rm if}~ z<0
\end{array}
\right.
.
\nn
\ee
The boundary conditions, at the surface $z=0$ and at $z\to \pm \infty$, are $\phi_{\bm q}(0^+)  = \phi_{\bm q}(0^-)$, $\phi_{\bm q}(z\to\pm \infty)  = 0$,  and
\be \label{eq:bc1}
\epsilon_{\rm b} \partial_z\phi_{\bm q}(z)\Big|_{z\to 0^+}  - \partial_z\phi_{\bm q}(z)\Big|_{z\to 0^-} = -4\pi e^2 
(n_{{\bm q},{\rm s}} + 1)
~.
\nn
\ee
Introducing the bulk Thomas-Fermi wavevector~\cite{Giuliani_and_Vignale} $q_{\rm TF}^2(q,\omega) =  -4\pi e^2 {\bar \chi}_{\rm b}({\bm q},\omega)/\epsilon_{\rm b}$, the solution of Eqs.~(\ref{eq:poisson_3}) with the boundary conditions above is
\be
\left\{
\begin{array}{ll}
\phi_{\bm q}(z) = {\bar \phi}_{{\bm q}} e^{-\sqrt{q^2+q_{\rm TF}^2(q,\omega)} z} & {\rm if}~ z>0
\vspace{0.2cm}\\
\phi_{\bm q}(z) = {\bar \phi}_{{\bm q}} e^{qz} & {\rm if}~ z<0
\end{array}
\right.
~,
\ee
where, using Eq.~(\ref{eq:bc1}), we get 
\be \label{eq:phi_bar_def}
{\bar \phi}_{{\bm q}} = \frac{4\pi e^2}{\epsilon \sqrt{q^2+q_{\rm TF}^2(q,\omega)} + q - 4\pi e^2 \chi_s({\bm q},\omega)}
~.
\ee
In the limit of zero frequency and small wavevectors, $\chi_{\rm s}({\bm q},\omega)$ becomes the densities of states of surface electrons.

\section{The cooling rate -- general theory}
We consider the kinetic equation for electrons at the surface of the topological insulator, which interact with those in the bulk via Coulomb interactions. 
The surface and bulk electrons are described by the Fermi distribution functions $f^{({\rm s})}_{{\bm k},\lambda}$ and $f^{({\rm b})}_{{\tilde {\bm k}},\eta}$ at the temperatures $T_{\rm s}$ and $T_{\rm b}$ and chemical potentials $\mu_{\rm s}$ and $\mu_{\rm b}$, respectively. As in Sect.~\ref{sect:model}, ${\bm k}$ and ${\tilde {\bm k}}$ are their two- and three-dimensional wavevectors, respectively. We recall that $\lambda = \pm 1$ is used to denote the two surface bands (together forming a Dirac cone), while $\eta = \pm 1$ is used for the bulk conduction and valence bands. 
For the calculation of the cooling time, we assume that no driving field is present, and that material parameters (including temperatures and chemical potentials) are isotropic. 
However, since the two populations are at different uniform chemical potentials and temperatures, we can study the time evolution of their distribution functions. The kinetic equation satisfied by $f^{({\rm s})}_{{\bm k},\lambda}$ is~\cite{Principi_prl_2017}
\begin{eqnarray} \label{eq:boltzmann_def}
&&
\partial_t f^{({\rm s})}_{{\bm k},\lambda} = - {\cal I}_{{\bm k},\lambda}^{(\rm sb)}
~,
\end{eqnarray}
where ${\cal I}_{{\bm k},\lambda}^{(\rm sb)}$ is the electron-electron collision integral between surface and bulk electrons [resulting from the interaction in Eq.~(\ref{eq:hee})]. This conserves their numbers separately, but allows for the exchange of energy between them. We will specify the collision integral in the following subsection. First, however, we will derive the general expression for the cooling rate.

To obtain the cooling rate, we multiply Eq.~(\ref{eq:boltzmann_def}) by the energy of the surface state, $\varepsilon^{({\rm s})}_{{\bm k},\lambda}$, and sum over all wavevectors ${\bm k}$ and all values of the surface-band index $\lambda$. The left-hand side of the so-obtained equation yields the time derivative of the energy stored in surface states, $\partial_t E_{\rm s}$. The latter is rewritten as $\partial_t E^{\rm s} = C_{\rm s} \partial_t T_s$, where the heat capacity of surface states is defined as~\cite{Principi_prl_2017}
\be \label{eq:surface_heat_cap}
C_{\rm s} = \sum_\lambda \int \frac{d^2{\bm k}}{(2\pi)^2} \xi^{({\rm s})}_{{\bm k},\lambda} \left(- \frac{\partial f^{({\rm s})}_{{\bm k},\lambda}}{\partial \xi^{({\rm s})}_{{\bm k},\lambda}}\right) 
\left[ \frac{\xi^{({\rm s})}_{{\bm k},\lambda}}{T_{\rm s}} + \frac{\partial \mu_{\rm s}}{\partial T_{\rm s}}\right].
\ee
Here, $\xi^{({\rm s})}_{{\bm k},\lambda} = \varepsilon^{({\rm s})}_{{\bm k},\lambda} - \mu_s$.
The derivative of the surface chemical potential $\mu_{\rm s} \equiv \mu_{\rm s}(T_{\rm s})$ with respect to temperature is obtained by imposing the conservation of the surface electron density~\cite{Principi_prl_2017}.
Defining  the power dissipated into bulk states as ${\cal Q} = \sum_{{\bm k},\lambda} {\cal I}_{{\bm k},\lambda}^{(\rm sb)}\varepsilon^{({\rm s})}_{{\bm k},\lambda} \equiv \gamma C_{\rm s} (T_{\rm s} - T_{\rm b})$, Eq.~(\ref{eq:boltzmann_def}) yields the cooling rate~\cite{Principi_prl_2017}
\be \label{eq:cooling_time_def}
\gamma = \frac{{\cal Q}}{(T_{\rm s} - T_{\rm b})C_{\rm s} }
~.
\ee
In the following we first define the collision integral and then calculate ${\cal Q}$, and thus $\gamma$.

\section{The cooling rate -- surface-to-bulk power dissipation}
To calculate the power dissipated into bulk states, we first have to obtain the collision integral due to the interaction of Eq.~(\ref{eq:hee}). 
Within the Fermi-golden-rule approximation, the collision integral on the right-hand side of Eq.~(\ref{eq:boltzmann_def}) reads
\begin{widetext}
\be \label{eq:I_ee_general}
{\cal I}_{{\bm k},\lambda}^{(\rm sb)} &=& 2\frac{2\pi}{\hbar {\cal A}^2 L_z} \sum_{{\bm k}', {\tilde {\bm k}}', {\bm q}} \sum_{\lambda'} \sum_{\eta,\eta'} \int_{-\infty}^\infty d\omega V_{{\bm q},{\bm k}',{\tilde {\bm k}}'}^2 F_{{\bm k},\lambda;{\bm k}+{\bm q},\lambda'} \delta(\varepsilon_{{\bm k},\lambda}^{({\rm s})} - \varepsilon_{{\bm k}+{\bm q},\lambda'}^{({\rm s})} + \omega) \delta(\varepsilon_{{\bm k}',\eta}^{({\rm b})} - \varepsilon_{{\tilde {\bm k}}',\eta'}^{({\rm b})} - \omega) \delta({\bm k}'_\parallel - {\tilde {\bm k}}'_\parallel -{\bm q})
\nonumber\\
&\times&
\big[ 
f_{{\bm k},\lambda}^{({\rm s})} f_{{\bm k}',\eta}^{({\rm b})} (1 - f_{{\bm k}+{\bm q},\lambda'}^{({\rm s})}) (1 - f_{{\tilde {\bm k}}',\eta'}^{({\rm b})}) 
- 
(1 - f_{{\bm k},\lambda}^{({\rm s})}) (1 - f_{{\bm k}',\eta}^{({\rm b})}) f_{{\bm k}+{\bm q},\lambda'}^{({\rm s})} f_{{\tilde {\bm k}}',\eta'}^{({\rm b})}
\big]
~,
\ee
\end{widetext}
where the factor $2$ upfront accounts for the spin degeneracy of bulk states, $\omega$ and ${\bm q}$ are the transferred energy and momentum parallel to the surface, ${\bm k}'_\parallel$ and ${\tilde {\bm k}}'_\parallel$ are the components of the three-dimensional momenta ${\bm k}'$ and ${\tilde {\bm k}}'$ parallel to the surface, ${\cal A}$ is the surface area of the topological insulator and $L_z$ its extension in the third dimension. In this equation, $F_{{\bm k},\lambda;{\bm k}+{\bm q},\lambda'}$ is the squared matrix element of the surface-electron density operator between incoming and outgoing scattering states. Since bulk states are standing waves, the $z$-components of the three-dimensional momenta ${\bm k}'$ and ${\tilde {\bm k}}'$ are taken to be positive. The matrix element of the screened Coulomb interaction $V_{{\bm q},{\bm k}',{\tilde {\bm k}}'}$ is obtained by integrating it over the incoming and outgoing scattering states, {\it i.e.}
\be \label{eq:coll_integral_matr_el_interaction}
V_{{\bm q},{\bm k}',{\tilde {\bm k}}'}^2 &=& 4\left| \int_0^\infty dz \sin(k_z' z) \sin({\tilde k}_z' z) e^{-\sqrt{q^2+q_{\rm TF}^2(q,\omega)} z} \right|^2
\nn
&\times&
\big| {\bar \phi}_{{\bm q}}  \big|^2
~.
\ee
This matrix element is manipulated in App.~\ref{app:manip_interaction} to give
\be \label{eq:coll_integral_matr_el_interaction_2}
V_{{\bm q},{\bm k}',{\tilde {\bm k}}'}^2 
&\simeq&
\big| {\bar \phi}_{{\bm q}}  \big|^2 \frac{\pi^2}{4 \sqrt{q^2+q_{\rm TF}^2(q,\omega)}} 
\delta(k_z' - {\tilde k}_z')
\nn
&\equiv& 
V_{\bm q}^2
\delta(k_z' - {\tilde k}_z')
~.
\ee
To obtain this expression, we have evaluated the integral on the right-hand side of Eq.~(\ref{eq:coll_integral_matr_el_interaction}) and approximated it in the limit of small $q$ and $q_{\rm TF}(q,\omega)$ [the latter is justified when the bulk is undoped and $k_{\rm B} T_{\rm b} \ll 2\Delta$]. In this case, the weight of the integral is located around the lines ${\tilde k}_z' = \pm k_z'$. Thus, we replaced Lorenzians of width $\sqrt{q^2+q_{\rm TF}^2(q,\omega)}$ with delta-functions. Finally, we used that ${\tilde k}_z'$ and $k_z'$ must be taken to be positive.
Putting Eq.~(\ref{eq:coll_integral_matr_el_interaction_2}) back into Eq.~(\ref{eq:I_ee_general}), after some lengthy but straightforward algebra we get
\be \label{eq:Q_result}
{\cal Q}
&=&
\frac{4}{\pi \hbar }
\int \frac{d^2{\bm q}}{(2\pi)^2} V_{\bm q}^2
 \int_{0}^\infty d\omega \omega 
\big[n^{({\rm s})}(\omega) - n^{({\rm b})}(\omega)\big] 
\nn
&\times&
\Im m \chi_{\rm b}({\bm q},\omega) \Im m \chi_{\rm s}({\bm q},\omega)
~,
\ee
where $n^{({\rm s/b})}(\omega) = \big[e^{\hbar \omega/(k_{\rm B} T_{\rm s/b})} - 1\big]^{-1}$. 
In Eq.~(\ref{eq:Q_result}),
\be
\Im m \chi_{\rm s}({\bm q},\omega) &=& -\frac{\pi}{{\cal A}}  \sum_{{\bm k},\lambda,\lambda'} ( f^{({\rm s})}_{{\bm k},\lambda} - f^{({\rm s})}_{{\bm k}+{\bm q},\lambda'}) F_{{\bm k},\lambda;{\bm k}+{\bm q},\lambda'} 
\nn
&\times&
\delta(\varepsilon_{{\bm k},\lambda}^{({\rm s})} - \varepsilon_{{\bm k}+{\bm q},\lambda'}^{({\rm s})} + \omega)
~,
\ee
and
\be \label{eq:density_density_bulk}
\Im m \chi_{\rm b}({\bm q},\omega) &=& - \frac{\pi}{{\cal A} L_z} \sum_{{\bm k}',\eta,\eta'} (f^{({\rm b})}_{{\bm k}',\eta} - f^{({\rm b})}_{{\bm k}'-{\bm q},\eta'}) 
\nn
&\times&
\delta(\varepsilon_{{\bm k}',\eta}^{({\rm b})} - \varepsilon_{{\bm k}' - {\bm q},\eta'}^{({\rm b})} - \omega)
~,
\ee
are the imaginary parts of the density-density response functions of surface~\cite{Wunsch_njp_2006,Barlas_prl_2007,Principi_prb_2009} and bulk~\cite{Giuliani_and_Vignale} states, respectively.
The function $\Im m \chi_{\rm b}({\bm q},\omega)$, not to be confused with the function ${\bar \chi}_{\rm b}({\bm q},\omega)$ introduced in Sect.~\ref{app:electrostatics}, is calculated in App.~\ref{app:bulk_response}. 

The cooling rate $\gamma$ is obtained by inserting ${\cal Q}$ in Eq.~(\ref{eq:Q_result}) back into the definition~(\ref{eq:cooling_time_def}), and reads
\be
\gamma &=& -\frac{4}{\pi \hbar C_s}
\int \frac{d^2{\bm q}}{(2\pi)^2} V_{\bm q}^2
\int_{0}^\infty d\omega \omega 
\frac{n^{({\rm s})}(\omega) - n^{({\rm b})}(\omega)}{T_s - T_{\rm b}} 
\nn
&\times&
\Im m \chi_{\rm b}({\bm q},\omega) \Im m \chi_{\rm s}({\bm q},\omega)
~.
\ee
This equation is the key result of this paper.

\begin{figure}[t]
\begin{center}
\begin{tabular}{c}
\begin{overpic}[width=1.0\columnwidth]{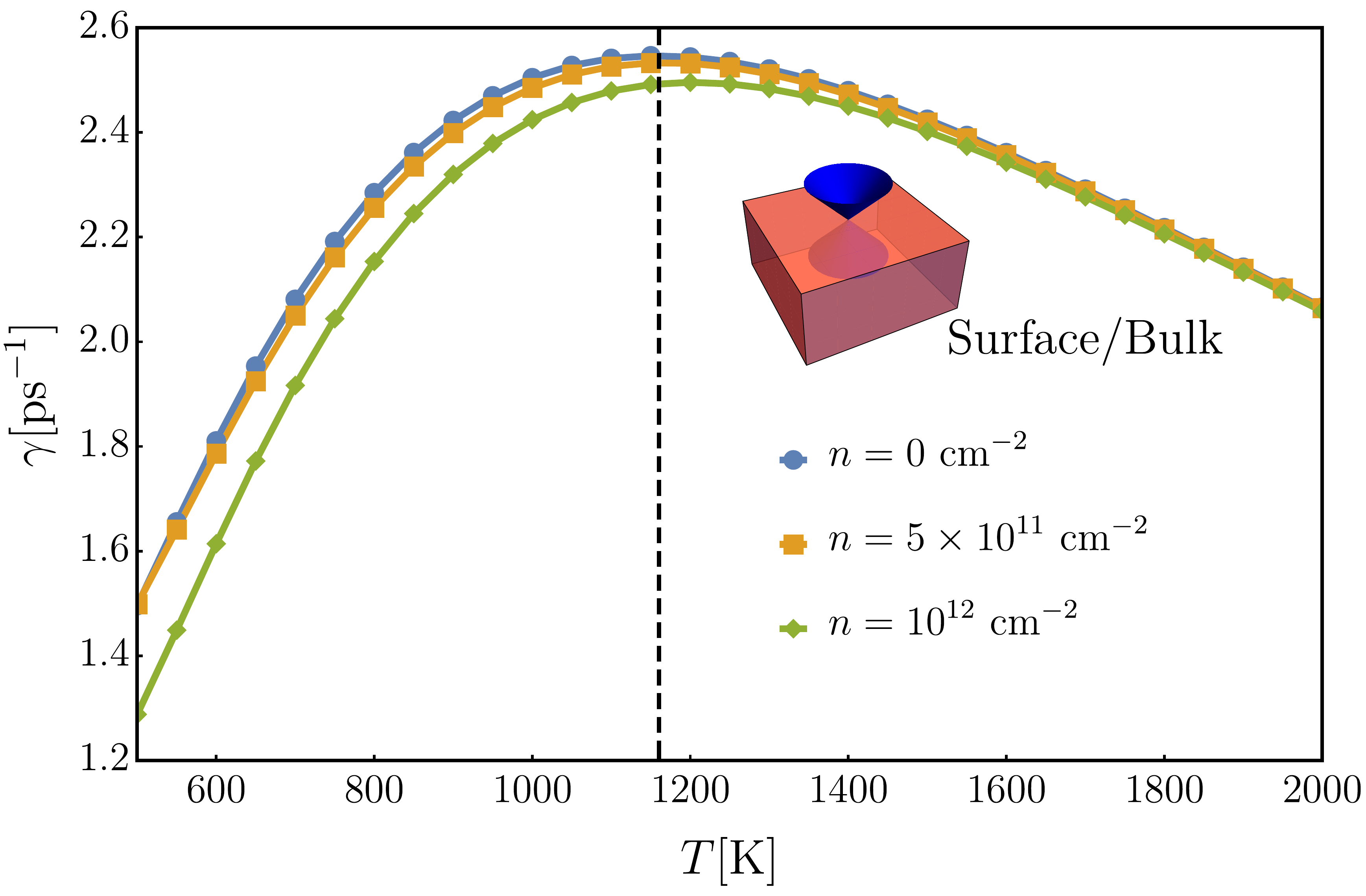}
\put(35,35){(a)}
\end{overpic}
\\
\begin{overpic}[width=1.0\columnwidth]{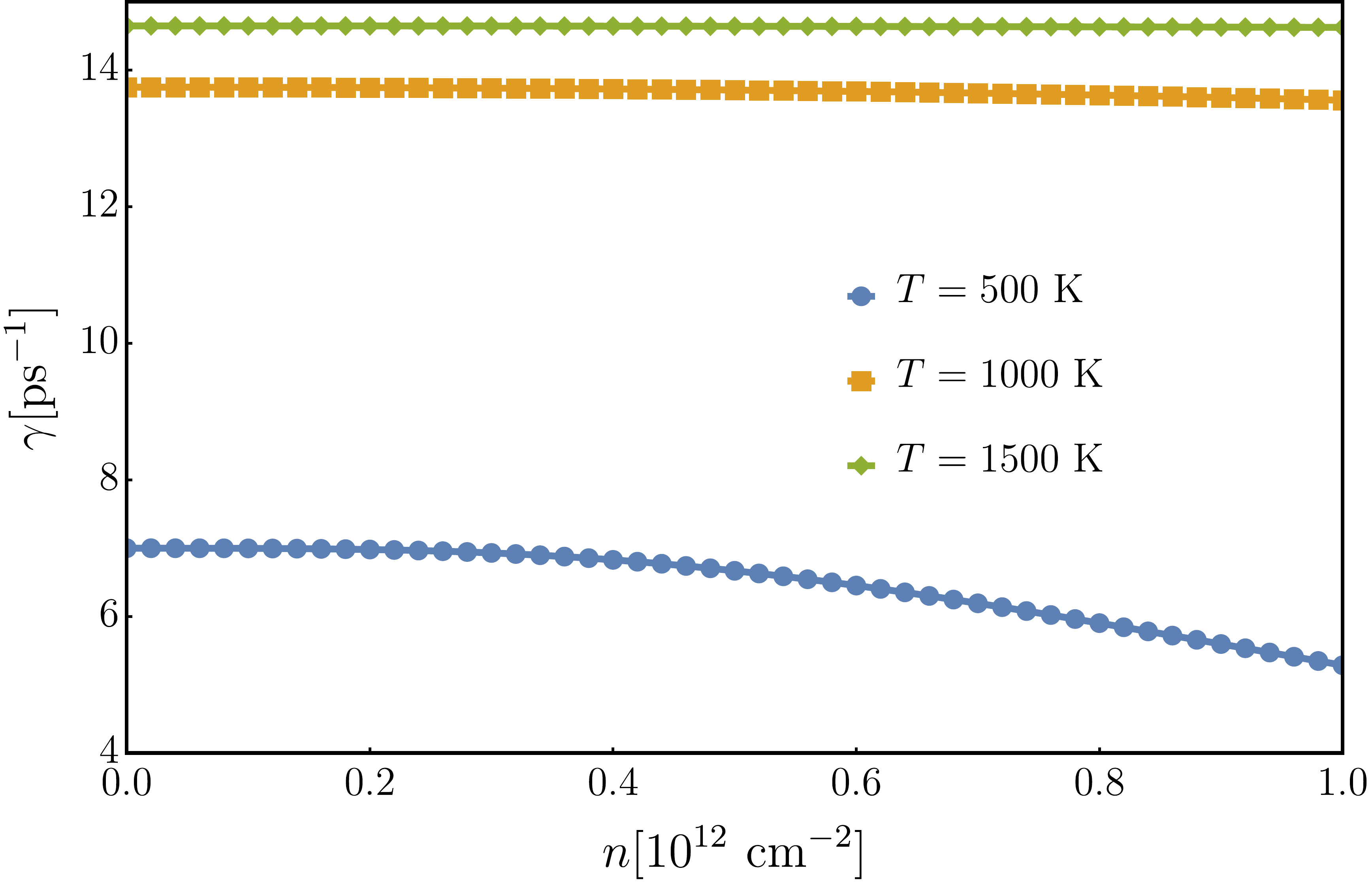}
\put(35,35){(b)}
\end{overpic}
\end{tabular}
\end{center}
\caption{
Panel (a): the cooling rate for Dirac-like surface states ($v_{\rm F} = 0.5 \times 10^6~{\rm m}/{\rm s}$) in proximity to an undoped 3D bulk ($n_b = 0$, which translates into $\mu_{\rm b} = 0$) kept at room temperature ($T_{\rm b} = 300~{\rm K}$, much smaller than the gap energy: $k_{\rm B}T_{\rm b}\ll 2\Delta$), calculated from Eq.~(\ref{eq:cooling_time_def}) and plotted as a function of temperature. 
Curves exhibit a maximum at a surface electron temperature corresponding to half the bulk-band gap, $\Delta = 100~{\rm meV}$ (dashed line). 
Panel (b): the cooling rate is very weakly dependent on surface carrier density.
The parameter used in these calculations are given at the beginning of Sect.~\ref{sect:results} and are recalled here for convenience. 
The bulk electron mass is set to~\cite{Orlita_prl_2015,Lang_acsnano_2012} $m=0.21m_e$, where $m_e = 9.1\times 10^{-31}~{\rm kg}$ is the bare electron mass, while the dielectric constant of the (undoped) topological insulator is taken to be~\cite{Kuznetsov_opteng_2021} $\epsilon_{\rm b} = 10$.
}
\label{fig:two}
\end{figure}

\section{Results}
\label{sect:results}
We now show our results for the cooling time of electrons in the surface state of a topological insulator and in graphene in proximity to a small-gap material. We find that the bulk behaves as an efficient heat sink for electrons, yielding sub-picosecond cooling times. 

We start with the surface states of the topological insulator. We numerically evaluate Eq.~(\ref{eq:cooling_time_def}), with the heat capacity of surface states and power lost to bulk states given by Eqs.~(\ref{eq:surface_heat_cap}) and~(\ref{eq:Q_result}), respectively. For a sake of definiteness, in the numerical calculations, we set the surface Fermi velocity~\cite{Tang_scirep_2013} $v_{\rm F} = 0.5 \times 10^6~{\rm m}/{\rm s}$, while the bulk is left undoped, {\it i.e.} its electron density is $n_b = 0$ (which translates into $\mu_{\rm b} = 0$). The bulk temperature is set to $T_{\rm b} = 300~{\rm K}$, which is much smaller than the bulk-band gap~\cite{Arakane_natcomm_2012} $2\Delta = 200~{\rm meV}$. Under these conditions, the bulk bands are nearly unpopulated and we can thus safely take $q_{\rm TF}(q,\omega) = 0$ in Eqs.~(\ref{eq:phi_bar_def}) and~(\ref{eq:coll_integral_matr_el_interaction}). Finally, the bulk electron mass is set to~\cite{Orlita_prl_2015,Lang_acsnano_2012} $m=0.21m_e$, where $m_e = 9.1\times 10^{-31}~{\rm kg}$ is the bare electron mass, while the dielectric constant of the (undoped) topological insulator is taken to be~\cite{Kuznetsov_opteng_2021} $\epsilon_{\rm b} = 10$. This value reflects the fact that the topological insulator we describe has no charge carriers in the bulk at low temperature (contrary to, e.g., common topological insulators such as ${\rm Bi}_2{\rm Se}_3$ which are metallic). The effect of bulk thermal excitations is included in our theory via the bulk response function. Furthermore, since the typical energies of excitations here are of the order of $k_{\rm B} T \sim 10-100~{\rm meV}$ (in particular, they can be interband transitions), the dielectric constant should be taken to be the one at intermediate frequencies, not the zero-frequency one.

In Fig.~\ref{fig:two} we show the cooling rate for surface-to-bulk Coulomb cooling in topological insulators (with Fermi energy in the bulk band gap). calculated from Eq.~(\ref{eq:cooling_time_def}). In Panel (a) we present our numerical results for three different values of the surface electronic density and as a function of temperature $T_{\rm s}$. The three chosen densities as $n_{\rm s} = 0~{\rm cm}^{-2}$, {\it i.e.} an undoped system with Fermi energy at the surface Dirac crossing, $n_{\rm s} = 5 \times 10^{11}~{\rm cm}^{-2}$ and $n_{\rm s} = 10^{12}~{\rm cm}^{-2}$. the latter corresponds to a Fermi energy close to the bottom of the bulk conduction band. We find that the cooling rate depends only weakly on surface carrier concentration, while it depends quite strongly on temperature. In particular, it decreases rapidly at low temperatures ({\it i.e.} for the surface temperature approaching the bulk one). Curiously, we find that the cooling rate exhibits a maximum at a temperature approximately equal to half the bulk-band gap. At this temperature, excitations of surface states occur with typical energies equal to $k_{\rm B} T_{\rm s}$, and become resonant with the bulk particle-hole excitations, whose weight grows sharply for energies larger than $2\Delta$. The latter fact is seen in the plots of the bulk density-density response (the bulk absorption spectrum) given in App.~\ref{app:bulk_response}.

\begin{figure}[t]
\begin{center}
\begin{tabular}{c}
\begin{overpic}[width=1.0\columnwidth]{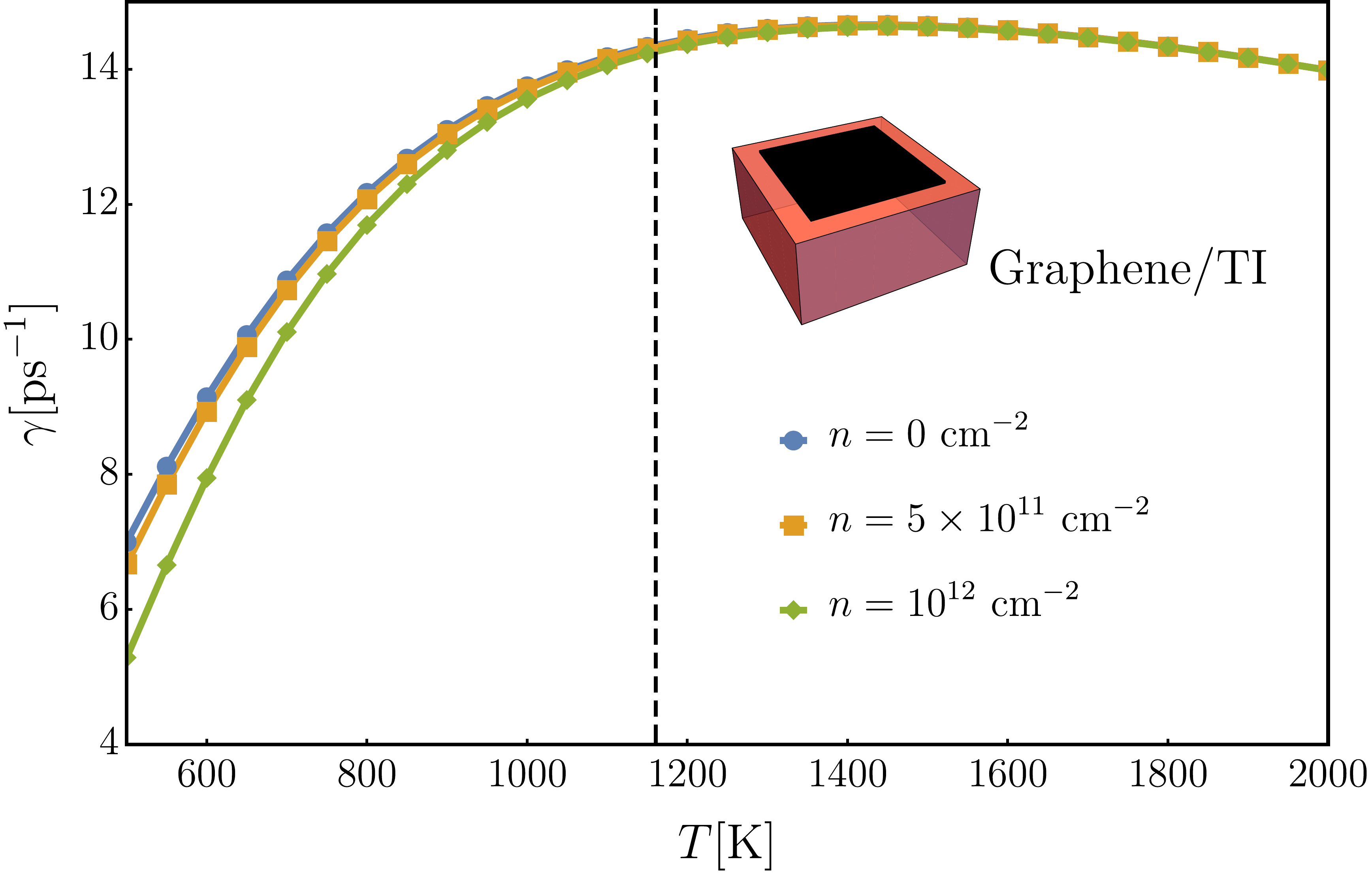}
\put(35,35){(a)}
\end{overpic}
\\
\begin{overpic}[width=1.0\columnwidth]{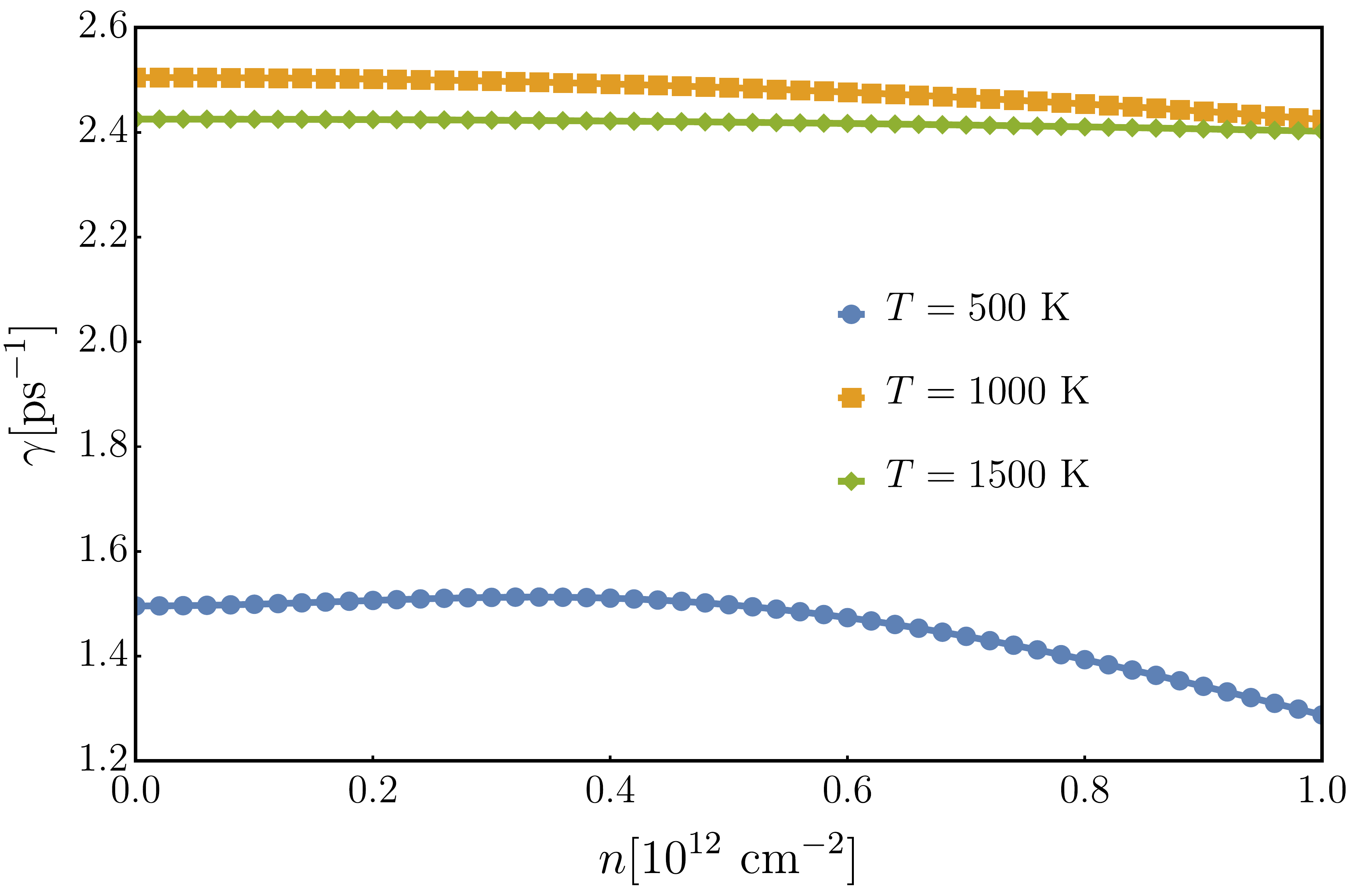}
\put(35,35){(b)}
\end{overpic}
\end{tabular}
\end{center}
\caption{
The cooling rate for Dirac-like states in graphene ($v_{\rm F} = 10^6~{\rm m}/{\rm s}$) in proximity to an undoped 3D bulk identical to that of Fig.~\ref{fig:two}, calculated from Eq.~(\ref{eq:cooling_time_def}).
Panel (a): as a function of temperature, it exhibits a maximum at a surface electron temperature corresponding to half the bulk-band gap $\Delta$ (dashed line). 
Panel (b): the cooling rate is very weakly dependent on surface carrier density.
The parameters used in these plots are given in Sect.~\ref{sect:results} and Fig.~\ref{fig:two}.
}
\label{fig:three}
\end{figure}

In Panel (b) we show the cooling rate as a function of the surface density $n_{\rm s}$ and for three values of the temperature, $T_{\rm s} = 500, 1000$ and $1500~{\rm K}$. We see that the numerical results are practically independent of carrier density at the highest temperatures, and only weakly dependent at the lowest one. This can be understood as a consequence of the complete smearing of the surface occupation function at temperatures much larger than the Fermi energy. Only at the lowest temperatures and highest densities achievable in our model we start observing some deviation from perfect flatness. 

In Fig.~\ref{fig:three} we show the cooling rate for a graphene sheet in proximity of a small-gap bulk material as a function of temperature [in panel (a)] and carrier density [in panel (b)]. For the bulk material, we use the same 3D topological insulator used above, {\it i.e.} the 3D parameters are taken to be the same. However, for the ``surface'' states (now a graphene sheet), the number of fermion flavors is set to~\cite{castroneto_rmp_2009} $N_{\rm f} = 4$, while the Fermi velocity is doubled, {\it i.e.} we use~\cite{castroneto_rmp_2009} $v_{\rm F}=10^6~{\rm m/s}$. Comparing Figs.~\ref{fig:two} and~\ref{fig:three}, we see that graphene would exhibit dynamics approximately four-five times faster than the topological-insulator surface states. This is surprising because, thanks to the doubling of the Fermi velocity and quadrupling of the number of fermion flavors, the density of states of undoped graphene and of the surface states studied above are (accidentally) identical. This in turn implies that, in the undoped limit, they also exhibit the same heat capacity.

The reason for the enhanced cooling rate is instead to be found in the typical energies of particle-hole excitations, which are different in the two systems. Due to the linear energy dispersion, the typical energy exchanged by surface and bulk states during a collision is $\omega \simeq \hbar v_{\rm F} q$. This in turn implies that, for a given value of momentum $\hbar q$, the energy exchanged between graphene and bulk electrons is twice the energy exchanged between surface and bulk states. 
Thus, the cooling dynamics proceeds at a faster pace in graphene, even if the number of interactions per unit time is the same as in a topological insulator, a fact that is reflected in the larger cooling rate.

\section{Discussion and conclusions}
In this paper we have developed the theory of the cooling dynamics of electrons in the surface states of a topological insulator, as well as in a graphene sheet, coupled to bulk states via non-retarded Coulomb interactions. The aim has been to explore the cooling capabilities of all-electronic surface-bulk coupling in these systems. For this reason, we have employed a simplified model: we have treated the surface states (and graphene electrons) as massless Dirac fermions characterized by a Dirac-like energy dispersion. We have neglected corrections due to, e.g., trigonal warping, which are expected to be only minor. We have considered a fully-gapped bulk with particle-hole symmetric parabolic-energy bands of equal masses. We have neglected all possible couplings between surface and bulk states (such as impurity or phonon-mediated hopping) with the exception of a density-density interaction of the Coulomb type. The latter conserves the number of electrons in surface and bulk states separately, but allows the exchange of energy between them. We have thus studied how such near-field radiative coupling can efficiently transfer energy between the surface and the bulk. We have thus explored an alternative mechanism to phonon-mediated cooling, which can be very efficient and could become dominant in some circumstances.

In fact, at least as a matter of principle, near-field radiative coupling could out-compete other cooling mechanisms. 
However, a comparison between different cooling pathways would require a significantly more detailed work to describe the impact of disorder, phonons, etc. on the cooling dynamics. All these mechanism are also potentially strongly dependent on material characteristics, and therefore it would be hard to derive general and universal trends. On the contrary, our model depends on few, experimentally available parameters (e.g., the surface Fermi velocity, the bulk-band mass, the undoped-material dielectric constant). Thus, it could be employed to study universal trends in electronic cooling of surfaces of topological insulators. 

We further observe that, although we have not treated phonon cooling in this paper, the interaction we describe could be thought as the result of phonon emission and re-absorption by the electrons at the surface and in the bulk, respectively. Such phonon-number-conserving processes could be easily incorporated in the theory and would result in a further enhancement of cooling rates. Finally, we stress that the theory also applies, with minor modifications, to graphene in proximity to narrow-gap materials, as we have shown above. This fact, which broadens the applicability of the present theory, has also important practical implications. Graphene is in fact one of the most studied materials for optoelectronic applications. The slowing down of cooling dynamics at high powers limits however its potential, for example to applications such as higher harmonic generation~\cite{Kovalev_npjqm_2021}. The fact that heat dissipation could be made more efficient via near-field coupling to narrow-gap materials offers a novel way to overcome the limitation intrinsic to current graphene devices.

\section{Acknowledgement}
A.P. acknowledges support from the European Commission under the EU Horizon 2020 MSCA-RISE-2019 programme (project 873028 HYDROTRONICS) and of the Leverhulme Trust under the grant RPG-2019-363. K.J.T. acknowledges funding from the European Union's Horizon 2020 research and innovation program under Grant Agreement No. 804349 (ERC StG CUHL).

\appendix

\renewcommand{\thefigure}{A\arabic{figure}}
\setcounter{figure}{0} 

\begin{widetext}

\section{Manipulation of the surface-bulk interaction}
\label{app:manip_interaction}
The integration of Eq.~(\ref{eq:coll_integral_matr_el_interaction}) yields
\be \label{eq:coll_integral_matr_el_interaction_app1}
V_{{\bm q},{\bm k}',{\tilde {\bm k}}'}^2
=
\big| {\bar \phi}_{{\bm q}}  \big|^2 
\left( 
\frac{\sqrt{q^2+q_{\rm TF}^2(q,\omega)}}{\big[q^2+q_{\rm TF}^2(q,\omega)\big] + (k_z' - {\tilde k}_z')^2} 
- 
\frac{\sqrt{q^2+q_{\rm TF}^2(q,\omega)}}{\big[q^2+q_{\rm TF}^2(q,\omega)\big] + (k_z' + {\tilde k}_z')^2} 
\right)^2
\ee
We now observe that, for small $q$ and $q_{\rm TF}^2(q,\omega)$, the regime of interest for the processes we are describing, the two terms on the right-hand side of Eq.~(\ref{eq:coll_integral_matr_el_interaction_app1}) are sharply peaked around ${\tilde k}_z' = k_z'$ and ${\tilde k}_z' = -k_z'$, respectively. Thus, we approximate them with two delta function, making sure that their total integral (over the variable ${\tilde k}_z'$) remains unchanged. We thus obtain
\be
V_{{\bm q},{\bm k}',{\tilde {\bm k}}'}^2
&\simeq&
\left(\frac{4\pi e^2}{(\epsilon_b +1) q - 4\pi e^2 \chi_s({\bm q},\omega)}\right)^2 \frac{\pi^2}{4 \sqrt{q^2+q_{\rm TF}^2(q,\omega)}} \big[\delta(k_z' - {\tilde k}_z') + \delta(k_z' + {\tilde k}_z')\big]
\nn
&\simeq&
\left(\frac{4\pi e^2}{(\epsilon_b +1) q - 4\pi e^2 \chi_s({\bm q},\omega)}\right)^2 \frac{\pi^2}{4 \sqrt{q^2+q_{\rm TF}^2(q,\omega)}} \delta(k_z' - {\tilde k}_z')
~.
\ee
In the last line we noticed that $k_z'$ and ${\tilde k}_z'$ must be taken as positive, since the bulk wavefunctions describe standing waves (negative wavevectors correspond to the same wavefunction).

\begin{figure}[t]
\begin{center}
\begin{tabular}{cc}
\begin{overpic}[width=0.49\columnwidth]{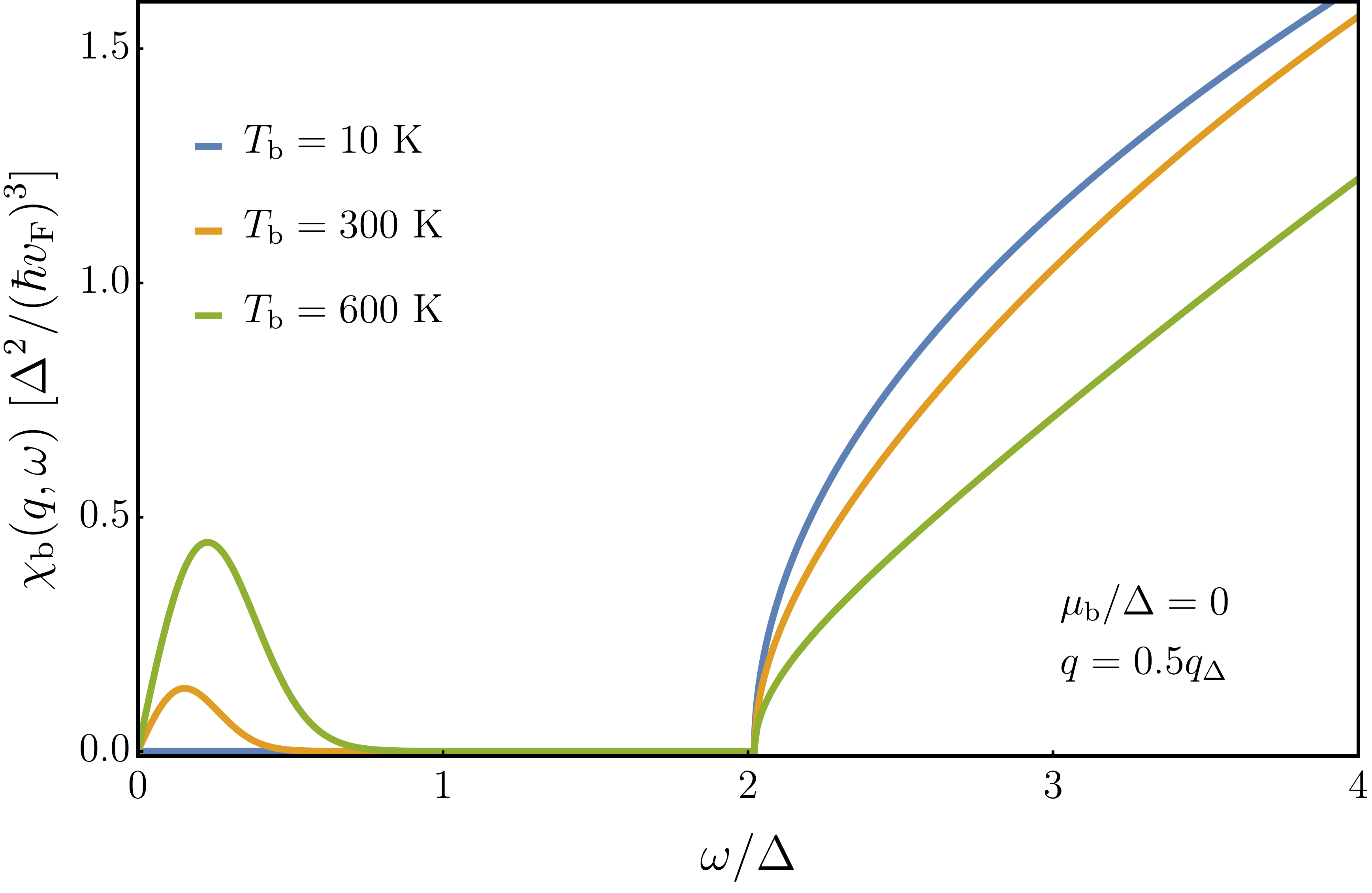}
\put(30,150){(a)}
\end{overpic}
&
\begin{overpic}[width=0.49\columnwidth]{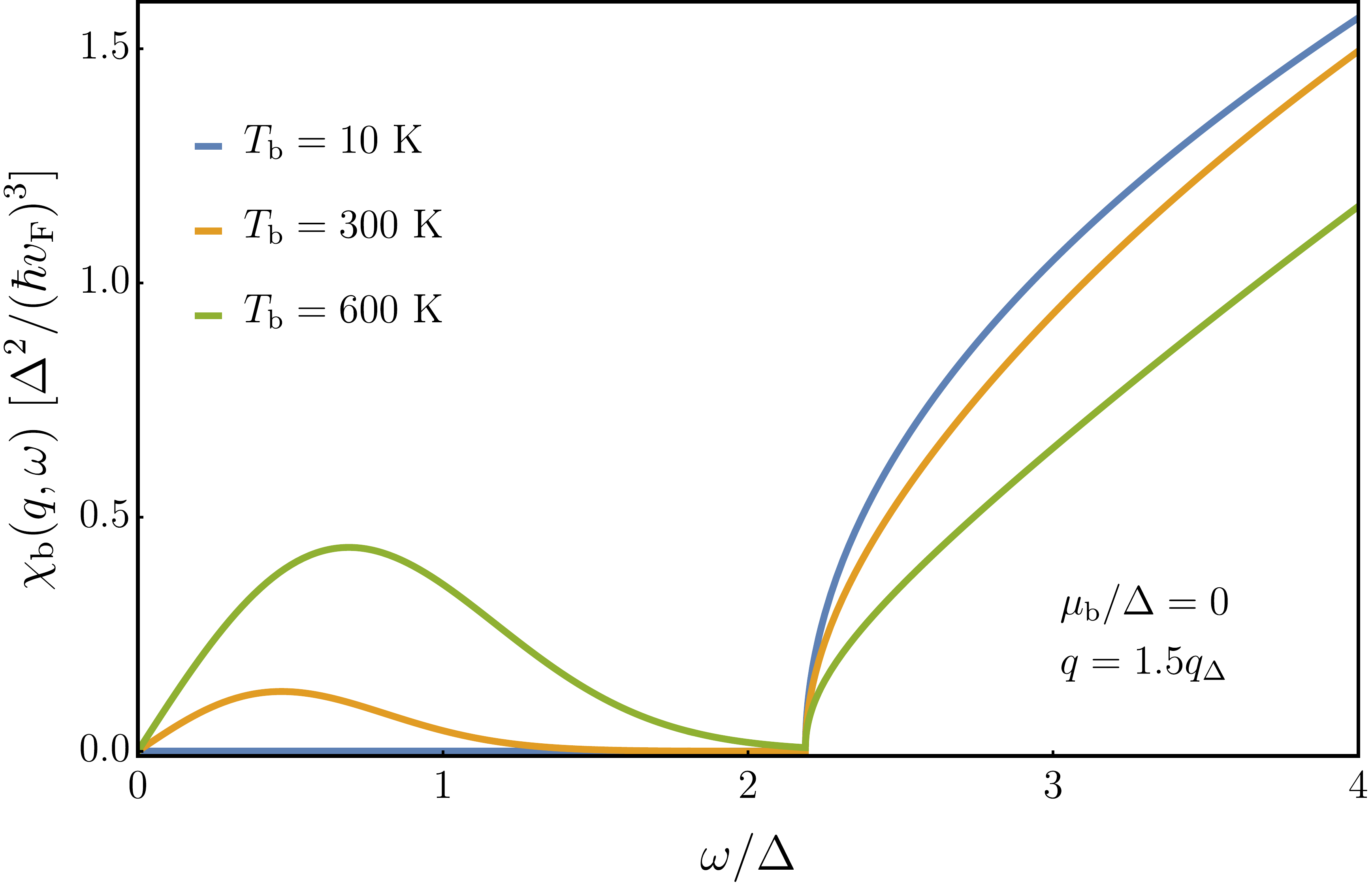}
\put(30,150){(b)}
\end{overpic}
\\
\begin{overpic}[width=0.49\columnwidth]{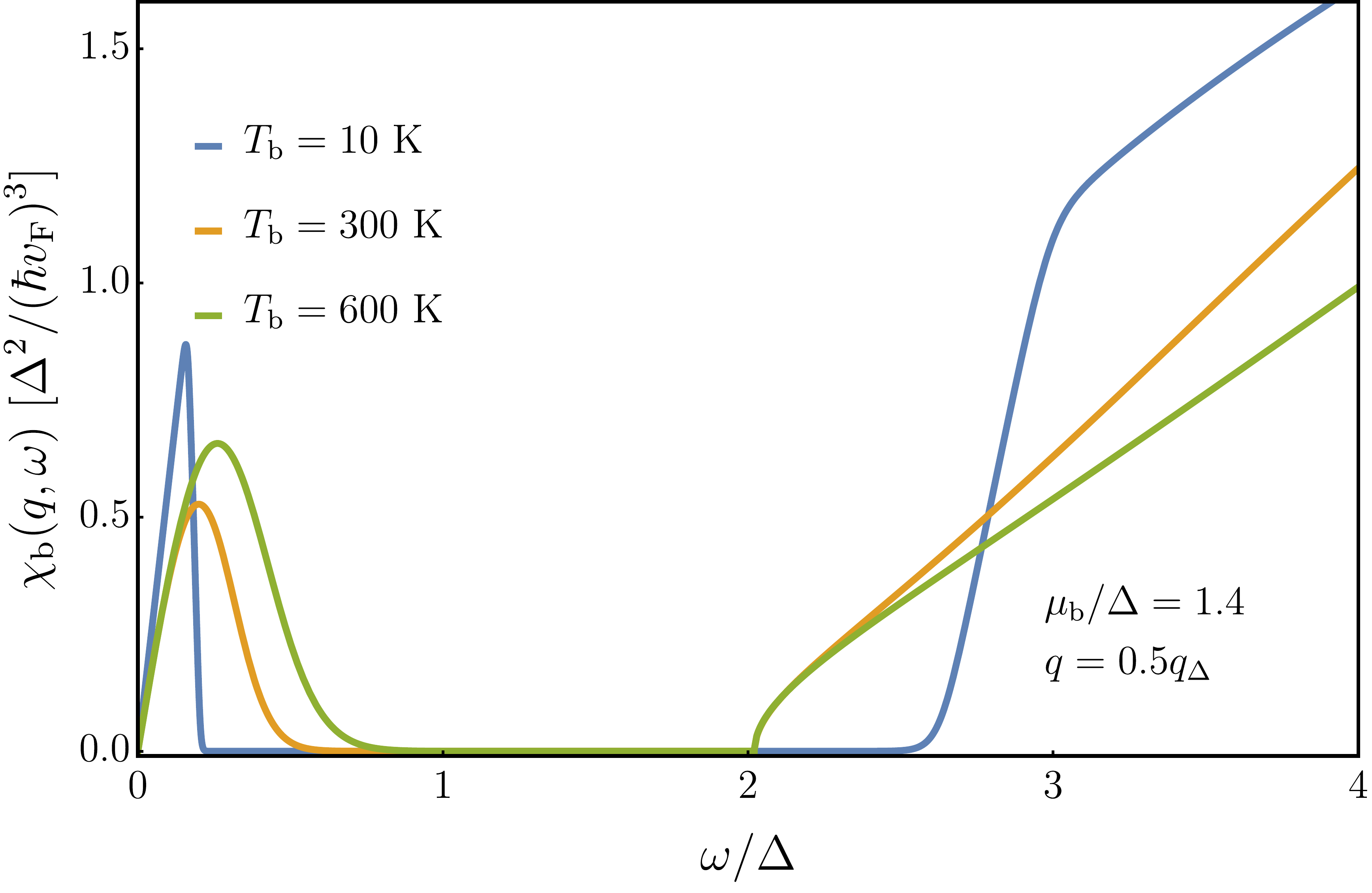}
\put(30,150){(c)}
\end{overpic}
&
\begin{overpic}[width=0.49\columnwidth]{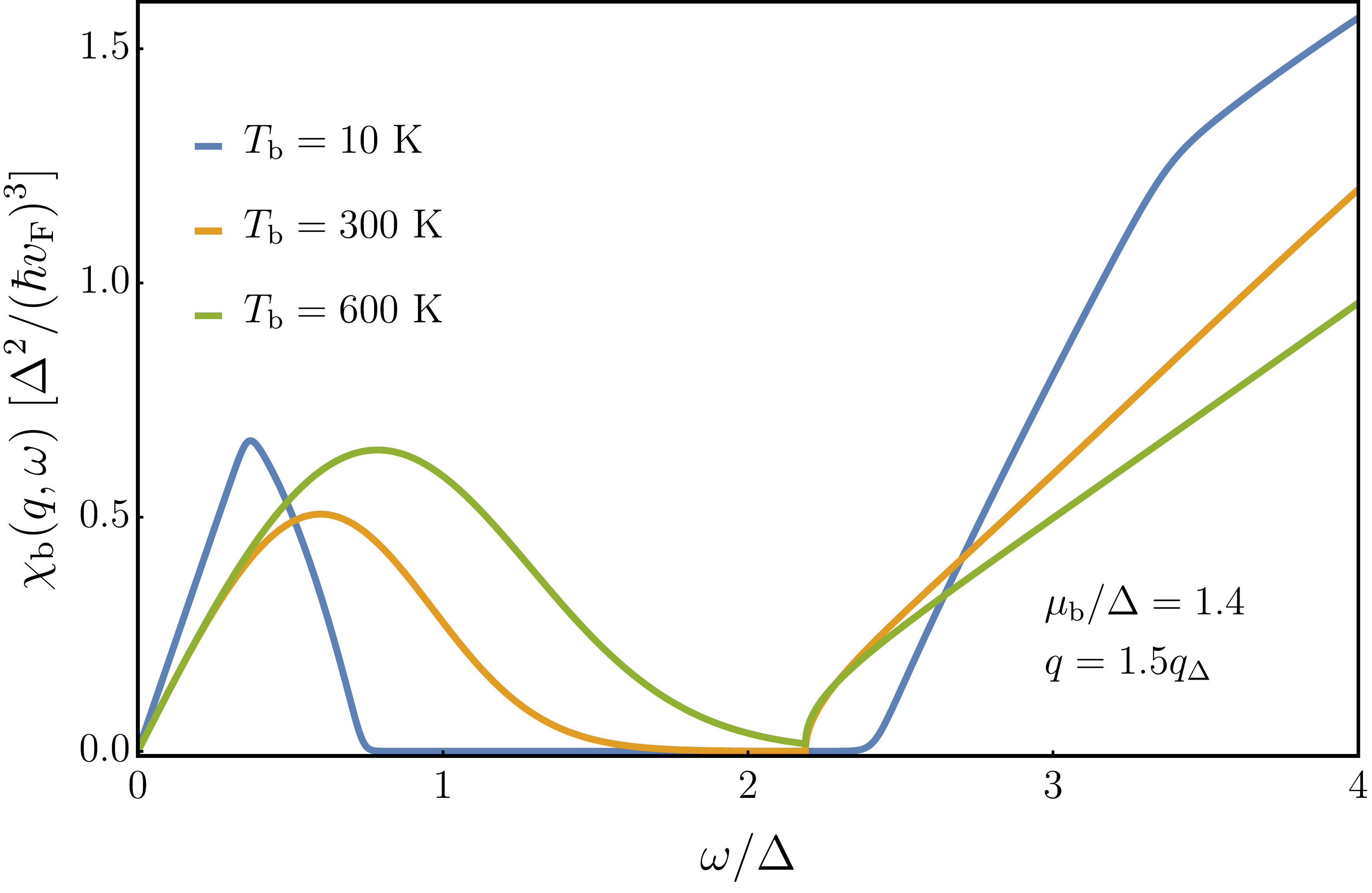}
\put(30,150){(d)}
\end{overpic}
\end{tabular}
\end{center}
\caption{
The density-density response function of bulk electrons in units of $\Delta^2/(\hbar v_{\rm F})^3$ ($v_{\rm F} = 0.5\times 10^6~{\rm m/s}$), for a fixed value of the wavevector $q$ and as a function of energy $\omega$ (in units of the half-gap $\Delta$). In each panel we show three curves, one for each temperature ($T= 10, 300$ and $600~{\rm K}$, respectively).
Panel (a): Here $q = 0.5 q_\Delta$ and the chemical potential is $\mu_{\rm b}=0$ (undoped system).
Panel (b): Same as in panel (a), but for $q = 1.5 q_\Delta$.
Panel (c): Here $q = 0.5 q_\Delta$ and the chemical potential is $\mu_{\rm b}=1.4\Delta$ (n-doped system).
Panel (d): Same as in panel (c), but for $q = 1.5 q_\Delta$.
The parameters used in these plots are the same as those used in Sect.~\ref{sect:results}, except that $\Delta = 50~{\rm meV}$. We also defined $q_\Delta = \Delta/(\hbar v_{\rm F})$.
}
\label{fig:supp_one}
\end{figure}

\section{The density-density response of bulk electrons}
\label{app:bulk_response}
The imaginary part of the density-density function in Eq.~(\ref{eq:density_density_bulk}) is given by
\be \label{eq:density_density_bulk_app}
\Im m\chi_b({\bm q},\omega) = -\pi\sum_{\eta,\eta'}\int \frac{d^3{\bm k}}{(2\pi)^3} \big[ f(\eta \varepsilon_{{\bm k}} + \eta \Delta - \mu_{\rm b}) - f(\eta \varepsilon_{{\bm k}} + \eta \Delta + \omega - \mu_{\rm b}) \big] \delta(\omega + \eta \varepsilon_{{\bm k}} + \eta \Delta - \eta' \varepsilon_{{\bm k}+{\bm q}} - \eta' \Delta)
~.
\nn
\ee
Here we rewrote $\varepsilon_{{\bm k},\eta} = \eta \Delta + \eta \varepsilon_{\bm k}$, introducing $\varepsilon_{\bm k} = k^2/(2m)$. We also defined $f(\xi) = \big[e^{\xi/(k_{\rm B} T_{\rm b})} + 1 \big]^{-1}$, and introduced the chemical potential of bulk bands $\mu_{\rm b}$. In what follows, we assume $\omega>0$ and analyze separately the intraband ($\eta' = \eta$) and interband ($\eta'\neq \eta$) contributions to Eq.~(\ref{eq:density_density_bulk_app}).

\subsection{Intraband term}
In this case, $\eta' = \eta$. Hence,
\be
\Im m\chi_b^{(\rm intra)}({\bm q},\omega) = -\frac{1}{4\pi} \sum_{\eta}\int_0^\infty dk k^2 \int_0^\pi d\theta \sin\theta \big[ f(\eta \varepsilon_{{\bm k}} + \eta \Delta - \mu_{\rm b}) - f(\eta \varepsilon_{{\bm k}} + \eta \Delta + \omega - \mu_{\rm b}) \big] \delta(\omega + \eta \varepsilon_{{\bm k}} - \eta \varepsilon_{{\bm k}+{\bm q}})
~.
\nn
\ee
The delta function implies that
\be
\omega + \eta \frac{k^2}{2m} - \eta \frac{k^2 + q^2 + 2 k q \cos(\theta_0)}{2m} = 0
\quad \Rightarrow \quad
\cos(\theta_0) = \frac{m}{k q} \left(\frac{q^2}{2m} - \eta \omega \right)
~.
\ee
Solutions exist for
\be
-\frac{k q}{m} \leq \frac{q^2}{2m} - \eta \omega \leq \frac{k q}{m}
\quad \Rightarrow \quad
k \geq \frac{m}{q} \left| \frac{q^2}{2m} - \eta \omega \right| \equiv k_0
\ee
Therefore,
\be \label{eq:im_chi_b_intra_final}
\Im m\chi_b^{(\rm intra)}({\bm q},\omega) &=& - \frac{1}{4\pi} \sum_{\eta}\int_0^\infty dk k^2 \big[ f(\eta \varepsilon_{{\bm k}} + \eta \Delta - \mu_{\rm b}) - f(\eta \varepsilon_{{\bm k}} + \eta \Delta + \omega - \mu_{\rm b}) \big] \frac{m}{k q} \int_0^\pi d\theta \delta\big(\theta - \theta_0\big)
\nn
&=&
- \frac{m}{4\pi q} \sum_{\eta}\int_{k_0}^\infty dk k \big[ f(\eta \varepsilon_{{\bm k}} + \eta \Delta - \mu_{\rm b}) - f(\eta \varepsilon_{{\bm k}} + \eta \Delta + \omega - \mu_{\rm b}) \big]
\nn
&=&
- \frac{m^2}{4\pi q} \sum_{\eta}\int_{k_0^2/(2m)}^\infty d\varepsilon \big[ f(\eta \varepsilon + \eta \Delta - \mu_{\rm b}) - f(\eta \varepsilon + \eta \Delta + \omega - \mu_{\rm b}) \big]
\nn
&=&
- \frac{m^2}{4\pi q} \sum_{\eta} \eta  \int_{k_0^2/(2m)}^\infty d\varepsilon \big[ f(\varepsilon + \Delta - \eta \mu_{\rm b}) - f(\varepsilon + \Delta + \eta \omega - \eta \mu_{\rm b}) \big]
\nn
&=&
- \frac{m^2 k_{\rm B} T}{4\pi q} \sum_{\eta} \eta \left\{ \ln\left[1 + \exp\left( \frac{\eta \mu_{\rm b} - \Delta}{k_{\rm B} T_{\rm b}} - \frac{k_0^2}{2m k_{\rm B} T_{\rm b}} \right) \right] - \ln\left[1 + \exp\left( \frac{\eta \mu_{\rm b} - \eta \omega - \Delta}{k_{\rm B} T_{\rm b}} - \frac{k_0^2}{2m k_{\rm B} T_{\rm b}} \right) \right] \right\}
~.
\nn
\ee

\subsection{Interband term}
In this case, $\eta' = -\eta$. Hence,
\be
\Im m\chi_b^{(\rm inter)}({\bm q},\omega) &=& -\frac{1}{4\pi} \sum_{\eta}\int_0^\infty dk k^2 \int_0^\pi d\theta \sin\theta  \big[ f(\eta \varepsilon_{{\bm k}} + \eta \Delta - \mu_{\rm b}) - f(\eta \varepsilon_{{\bm k}} + \eta \Delta + \omega - \mu_{\rm b}) \big] 
\nn
&\times&
\delta(\omega + 2\eta \Delta + \eta \varepsilon_{{\bm k}} + \eta \varepsilon_{{\bm k}+{\bm q}})
~.
\nn
\ee
It is clear that, for the delta function not to vanish, it must be $\eta = - 1$.
This in turn implies that
\be
\omega - 2 \Delta - \frac{k^2}{2m} - \frac{k^2 + q^2 + 2 k q \cos(\theta_0)}{2m} = 0
\quad \Rightarrow \quad
\cos(\theta_0) = \frac{m}{k q} \left(\omega - 2 \Delta - \frac{2 k^2 + q^2}{2m}  \right)
~.
\ee
Solutions exist for
\be
-\frac{k q}{m} \leq \omega - 2 \Delta - \frac{2 k^2 + q^2}{2m} \leq \frac{k q}{m}
\quad \Rightarrow \quad
\left\{
\begin{array}{l}
{\displaystyle k^2 - k q +  2 m \Delta - m \omega + \frac{q^2}{2} \leq 0 }
\vspace{0.2cm}\nn
{\displaystyle k^2 + k q +  2 m \Delta - m \omega + \frac{q^2}{2} \geq 0 }
\end{array}
\right.
\quad \Rightarrow \quad
|k_-| < k < k_+ 
~,
\ee
where
\be
k_\pm = \frac{q \pm \sqrt{4 (m \omega-2 m \Delta) - 2q^2}}{2}
~,
~~
\omega > 2\Delta + \frac{q^2}{2m}
~.
\ee
Therefore,
\be \label{eq:im_chi_b_inter_final}
\Im m\chi_b^{(\rm inter)}({\bm q},\omega) &=& -\frac{1}{4\pi} \int_0^\infty dk k^2 \big[ f(- \varepsilon_{{\bm k}} - \Delta - \mu_{\rm b}) - f(- \varepsilon_{{\bm k}} - \Delta + \omega - \mu_{\rm b}) \big] \frac{m}{k q} \int_0^\pi d\theta \delta\big(\theta - \theta_0\big)
\nn
&=&
- \frac{m}{4\pi q} \int_{|k_-|}^{k_+} dk k \big[ f(- \varepsilon_{{\bm k}} - \Delta - \mu_{\rm b}) - f(- \varepsilon_{{\bm k}} - \Delta + \omega - \mu_{\rm b}) \big]
\nn
&=&
\frac{m^2}{4\pi q} \int_{k_-^2/(2m)}^{k_+^2/(2m)} d\varepsilon \big[ f(\varepsilon + \Delta + \mu_{\rm b}) - f(\varepsilon + \Delta - \omega + \mu_{\rm b}) \big]
\nn
&=&
\frac{m^2 k_{\rm B} T}{4\pi q} \left\{ \ln\left[1 + \exp\left( -\frac{\mu_{\rm b} + \Delta}{k_{\rm B} T_{\rm b}} - \frac{k_-^2}{2m k_{\rm B} T_{\rm b}} \right) \right] - \ln\left[1 + \exp\left( -\frac{\mu_{\rm b} + \Delta}{k_{\rm B} T_{\rm b}} - \frac{k_+^2}{2m k_{\rm B} T_{\rm b}} \right) \right] 
\right.
\nn
&-&
\left.
\ln\left[1 + \exp\left( -\frac{\mu_{\rm b} - \omega + \Delta}{k_{\rm B} T_{\rm b}} - \frac{k_-^2}{2m k_{\rm B} T_{\rm b}} \right) \right] + \ln\left[1 + \exp\left( -\frac{\mu_{\rm b} - \omega + \Delta}{k_{\rm B} T_{\rm b}} - \frac{k_+^2}{2m k_{\rm B} T_{\rm b}} \right) \right] \right\}
~.
\ee

\subsection{Results}
Fig.~\ref{fig:supp_one} shows results for the imaginary part of the bulk density-density response function, $\Im m\chi_b({\bm q},\omega)$, where its intra- and inter-band parts are given in Eqs.~(\ref{eq:im_chi_b_intra_final}) and~(\ref{eq:im_chi_b_inter_final}), respectively. The parameters used in these plots are those given in Sect.~\ref{sect:results}, with the exception of the half-gap which has been set to $\Delta = 50~{\rm meV}$ for convenience reasons. 
Different panels correspond to different values of the chemical potential and of the wavevector $q$ [in units of $q_\Delta = \Delta/(\hbar v_{\rm F})$] used.
In each plot the density-density response function of bulk electrons is shows in units of $\Delta^2/(\hbar v_{\rm F})^3$, where $v_{\rm F} = 0.5 \times 10^6 ~{\rm m/s}$,
as a function of energy $\omega$. The latter is in units of the half-gap $\Delta$. In each panel we show three curves, one for each temperature, $T= 10, 300$ and $600~{\rm K}$, respectively.

\end{widetext}

\bibliography{refs}

\end{document}